\documentclass{article}
\usepackage{amssymb}
\usepackage{amsfonts}
\usepackage{amsmath}

\begin{document}

\title{Supersymmetric content of the Dirac and Duffin-Kemmer-Petiau equations%
}
\author{A. Okninski \\
Physics Division, Politechnika Swietokrzyska, Al. 1000-lecia PP 7, \\
25-314 Kielce, Poland}
\maketitle

\begin{abstract}
We study subsolutions of the Dirac and Duffin-Kemmer-Petiau equations
described in our earlier papers. It is shown that subsolutions of the
Duffin-Kemmer-Petiau equations and those of the Dirac equation obey the same
Dirac equation with some built-in projection operator. This covariant
equation can be referred to as supersymmetric since it has bosonic as well
as fermionic degrees of freedom.
\end{abstract}

\section{Introduction}

Recently, subsolutions of the Duffin-Kemmer-Petiau (DKP) equations were
found and it was shown that the subsolutions fulfill the appropriately
projected Dirac equation \cite{Okninski2003,Okninski2004}. On the other
hand, massive subsolutions of the Dirac equation were also found and studied 
\cite{Okninski2007}. In the present paper we demonstrate that subsolutions
of the DKP equations and those of the Dirac equation obey the same Dirac
equation with some built-in projection operator. This equation was shown to
be covariant in our earlier paper \cite{Okninski2007}. We shall refer to
this equation as supersymmetric since it has bosonic (spin $0$ and $1$) as
well as fermionic (spin $\frac{1}{2}$) degrees of freedom.

Some of the results described below were derived earlier but are included
for the sake of completeness. The paper is organized as follows. In Section
2 the Dirac as well as the DKP equations are described shortly. The DKP
equations for $s=0$ are written as a set of two $3\times 3$ equations in
Section 3 (the case $s=1$ leads to analogous equations \cite{Okninski2003})
and it is shown that their solutions fulfill the Dirac equation. In Section
4 subsolutions of the Dirac equation are described. Then in Section 5 it is
demonstrated that all these subsolutions obey the Dirac equation with
built-in projection operator, referred henceforth as the supersymmetric
equation. Finally, the supersymmetric equation is written in representation
independent form (its covariance was verified in \cite{Okninski2007}).

\section{Relativistic equations}

In what follows tensor indices are denoted with Greek letters, $\mu =0,1,2,3$%
. We shall use the following convention for the Minkowski space-time metric
tensor: $g^{\mu \nu }=$ \textrm{diag}$\left( 1,-1,-1,-1\right) $ and we
shall always sum over repeated indices. Four-momentum operators are defined
in natural units, $c=1$, $\hslash =1$, as $p^{\mu }=i\frac{\partial }{%
\partial x_{\mu }}$.

\subsection{Dirac equation}

The Dirac equation is a relativistic quantum mechanical wave equation
formulated by Paul Dirac in 1928 providing a description of elementary spin-$%
\frac{1}{2}$ particles, such as electrons and quarks, consistent with both
the principles of quantum mechanics and the theory of special relativity 
\cite{Dirac1928}. The Dirac equation is \cite{Bjorken1964, Berestetskii1971,
Thaller1992}:

\begin{equation}
\gamma ^{\mu }p_{\mu }\Psi =m\Psi ,  \label{Dirac1}
\end{equation}%
where $m$ is the rest mass of the elementary particle. The $\gamma $'s are $%
4\times 4$ anticommuting Dirac matrices: $\gamma ^{\mu }\gamma ^{\nu
}+\gamma ^{\nu }\gamma ^{\mu }=2g^{\mu \nu }I$ where $I$ is a unit matrix.
In the spinor representation of the Dirac matrices we have $\gamma
^{0}=\left( 
\begin{array}{cc}
\mathbf{0} & \sigma ^{0} \\ 
\sigma ^{0} & \mathbf{0}%
\end{array}%
\right) $, $\gamma ^{j}=\left( 
\begin{array}{cc}
\mathbf{0} & -\sigma ^{j} \\ 
\sigma ^{j} & \mathbf{0}%
\end{array}%
\right) $, $j=1,2,3$, $\gamma ^{5}=\left( 
\begin{array}{cc}
\sigma ^{0} & \mathbf{0} \\ 
\mathbf{0} & -\sigma ^{0}%
\end{array}%
\right) $. The wave function is a bispinor, i.e. consists of $2$
two-component spinors $\xi $, $\eta $: $\Psi =\left( 
\begin{array}{c}
\xi \\ 
\eta%
\end{array}%
\right) $.

\subsection{Duffin-Kemmer-Petiau equations}

The DKP equations for spin $0$ and $1$ are written as:

\begin{equation}
\beta _{\mu }p^{\mu }\Psi =m\Psi ,  \label{KDP-s0,1}
\end{equation}%
with $5\times 5$ and $10\times 10$ matrices $\beta ^{\mu }$, respectively,
which fulfill the following commutation relations \cite{Duffin1938,
Kemmer1939}:%
\begin{equation}
\beta ^{\lambda }\beta ^{\mu }\beta ^{\nu }+\beta ^{\nu }\beta ^{\mu }\beta
^{\lambda }=g^{\lambda \mu }\beta ^{\nu }+g^{\nu \mu }\beta ^{\lambda }.
\label{algebra-b}
\end{equation}

In the case of $5\times5$ (spin-$0$) representation of $\beta^{\mu}$
matrices Eq.(\ref{KDP-s0,1}) is equivalent to the following set of
equations: 
\begin{equation}
\left. 
\begin{array}{ccc}
p^{\mu}\psi & = & m\psi^{\mu} \\ 
p_{\nu}\psi^{\nu} & = & m\psi%
\end{array}
\right\} ,  \label{KDP-s0-1}
\end{equation}
if we define $\Psi$ in (\ref{KDP-s0,1}) as:

\begin{equation}
\Psi=\left( \psi^{\mu},\psi\right) ^{T}=\left( \psi^{0},\psi^{1},\psi
^{2},\psi^{3},\psi\right) ^{T},  \label{wavef-0}
\end{equation}
where $^{T}$ denotes transposition of a matrix. Let us note that Eq.(\ref%
{KDP-s0-1}) can be obtained by factorizing second-order derivatives in the
Klein-Gordon equation $p_{\mu}p^{\mu}\,\psi=m^{2}\psi$.

In the case of $10\times10$ (spin-$1$) representation of matrices $%
\beta^{\mu }$ Eq.(\ref{KDP-s0,1}) reduces to:

\begin{equation}
\left. 
\begin{array}{ccc}
p^{\mu }\psi ^{\nu }-p^{\nu }\psi ^{\mu } & = & m\psi ^{\mu \nu } \\ 
p_{\mu }\psi ^{\mu \nu } & = & m\psi ^{\nu }%
\end{array}%
\right\} ,  \label{KDP-s1-1}
\end{equation}%
with the following definition of $\Psi $ in (\ref{KDP-s0,1}): 
\begin{equation}
\Psi =\left( \psi ^{\mu \nu },\psi ^{\lambda }\right) ^{T}=\left( \psi
^{01},\psi ^{02},\psi ^{03},\psi ^{23},\psi ^{31},\psi ^{12},\psi ^{0},\psi
^{1},\psi ^{2},\psi ^{3}\right) ^{T},  \label{wavef-1}
\end{equation}%
where $\psi ^{\lambda }$ are real and $\psi ^{\mu \nu }$ are purely
imaginary (in alternative formulation we have $-\partial ^{\mu }\psi ^{\nu
}+\partial ^{\nu }\psi ^{\mu }=m\psi ^{\mu \nu }$, $\partial _{\mu }\psi
^{\mu \nu }=m\psi ^{\nu }$, where $\psi ^{\lambda }$, $\psi ^{\mu \nu }$ are
real). Because of antisymmetry of $\psi ^{\mu \nu }$ we have $p_{\nu }\psi
^{\nu }=0$ what implies spin $1$ condition. The set of equations (\ref%
{KDP-s1-1}) was first written by Proca \cite{Proca1936} and in a different
context by Lanczos \cite{Lanczos1929}. More on the rich history of the
formalism of Duffin, Kemmer and Petiau can be found in \cite{Bogush2007}.

\section{Splitting the spin-$0$ Duffin-Kemmer-Petiau equations}

Four-vectors $\psi ^{\mu }=\left( \psi ^{0},\mathbf{\psi }\right) $ and
spinors $\zeta ^{A\dot{B}}$ are related by formula: 
\begin{equation}
\zeta ^{A\dot{B}}=\left( \sigma ^{0}\psi ^{0}+\mathbf{\sigma }\cdot \mathbf{%
\psi }\right) ^{A\dot{B}}=\left( 
\begin{array}{cc}
\zeta ^{1\dot{1}} & \zeta ^{1\dot{2}} \\ 
\zeta ^{2\dot{1}} & \zeta ^{2\dot{2}}%
\end{array}%
\right) =\left( 
\begin{array}{cc}
\psi ^{0}+\psi ^{3} & \psi ^{1}-i\psi ^{2} \\ 
\psi ^{1}+i\psi ^{2} & \psi ^{0}-\psi ^{3}%
\end{array}%
\right) ,  \label{4vector-spinor}
\end{equation}%
where $A,\dot{B}$ number rows and columns, respectively, and $\sigma ^{j}$, $%
j=1,2,3$, are the Pauli matrices, $\sigma ^{0}$\ is the unit matrix. For
details of the spinor calculus reader should consult \cite%
{Berestetskii1971,MTW1973,Corson1953}.

Equations (\ref{KDP-s0-1}) can be written within spinor formalism as:

\begin{equation}
\left. 
\begin{array}{ccc}
p^{A\dot{B}}\psi & = & m\psi ^{A\dot{B}} \\ 
p_{A\dot{B}}\psi ^{A\dot{B}} & = & 2m\psi%
\end{array}%
\right\} .  \label{KDP-s0-2}
\end{equation}

It follows from (\ref{KDP-s0-2}) that $p_{A\dot{B}}\psi ^{A\dot{B}}=p_{A\dot{%
B}}p^{A\dot{B}}\psi $ and $p_{A\dot{B}}p^{A\dot{B}}\psi =2m^{2}\psi $.
Moreover, $p_{A\dot{B}}p^{A\dot{B}}=p_{1\dot{1}}p^{1\dot{1}}+p_{2\dot{1}}p^{2%
\dot{1}}+p_{1\dot{2}}p^{1\dot{2}}+p_{2\dot{2}}p^{2\dot{2}}=2p_{\mu }p^{\mu }$
and the Klein-Gordon equation $p_{\mu }p^{\mu }\psi =m^{2}\psi $ follows.
Let us note that due to spinor identities $p_{1\dot{1}}p^{1\dot{1}}+p_{2\dot{%
1}}p^{2\dot{1}}=p_{\mu }p^{\mu }$, $p_{1\dot{2}}p^{1\dot{2}}+p_{2\dot{2}}p^{2%
\dot{2}}=p_{\mu }p^{\mu }$ we can split the last of equations (\ref{KDP-s0-2}%
) and write Eqs.(\ref{KDP-s0-2}) as a set of two equations:%
\begin{equation}
\left. 
\begin{array}{r}
p^{1\dot{1}}\psi =m\psi ^{1\dot{1}} \\ 
p^{2\dot{1}}\psi =m\psi ^{2\dot{1}} \\ 
p_{1\dot{1}}\psi ^{1\dot{1}}+p_{2\dot{1}}\psi ^{2\dot{1}}=m\psi%
\end{array}%
\right\} ,  \label{const-s0-1}
\end{equation}%
\begin{equation}
\left. 
\begin{array}{r}
p^{1\dot{2}}\psi =m\psi ^{1\dot{2}} \\ 
p^{2\dot{2}}\psi =m\psi ^{2\dot{2}} \\ 
p_{1\dot{2}}\psi ^{1\dot{2}}+p_{2\dot{2}}\psi ^{2\dot{2}}=m\psi%
\end{array}%
\right\} ,  \label{const-s0-2}
\end{equation}%
each of which describes particle with mass $m$ (we check this substituting
e.g. $\psi ^{1\dot{1}}$, $\psi ^{2\dot{1}}$ or $\psi ^{1\dot{2}}$, $\psi ^{2%
\dot{2}}$\ into the third equations). Eq. (\ref{KDP-s0-2}) and the set of
two equations (\ref{const-s0-1}), (\ref{const-s0-2}) are equivalent. We
described equations (\ref{const-s0-1}), (\ref{const-s0-2}) in \cite%
{Okninski1981,Okninski1982}. From each of equations(\ref{const-s0-1}), (\ref%
{const-s0-2}) an identity follows:%
\begin{eqnarray}
p^{2\dot{1}}\psi ^{1\dot{1}} &=&p^{1\dot{1}}\psi ^{2\dot{1}},
\label{identities0-a} \\
p^{2\dot{2}}\psi ^{1\dot{2}} &=&p^{1\dot{2}}\psi ^{2\dot{2}}.
\label{identities0-b}
\end{eqnarray}

Equation (\ref{const-s0-1}) and the identity (\ref{identities0-a}), as well
as equation (\ref{const-s0-2}) and the identity (\ref{identities0-b}) can be
written in form of the Dirac equations:%
\begin{equation}
\left( 
\begin{array}{cccc}
0 & 0 & p^{0}+p^{3} & p^{1}-ip^{2} \\ 
0 & 0 & p^{1}+ip^{2} & p^{0}-p^{3} \\ 
p^{0}-p^{3} & -p^{1}+ip^{2} & 0 & 0 \\ 
-p^{1}-ip^{2} & p^{0}+p^{3} & 0 & 0%
\end{array}%
\right) \left( 
\begin{array}{c}
\psi ^{1\dot{1}} \\ 
\psi ^{2\dot{1}} \\ 
\chi \\ 
0%
\end{array}%
\right) =m\left( 
\begin{array}{c}
\psi ^{1\dot{1}} \\ 
\psi ^{2\dot{1}} \\ 
\chi \\ 
0%
\end{array}%
\right) ,  \label{A-DKP}
\end{equation}%
\begin{equation}
\left( 
\begin{array}{cccc}
0 & 0 & p^{0}-p^{3} & p^{1}+ip^{2} \\ 
0 & 0 & p^{1}-ip^{2} & p^{0}+p^{3} \\ 
p^{0}+p^{3} & -p^{1}-ip^{2} & 0 & 0 \\ 
-p^{1}+ip^{2} & p^{0}-p^{3} & 0 & 0%
\end{array}%
\right) \left( 
\begin{array}{c}
\psi ^{2\dot{2}} \\ 
\psi ^{1\dot{2}} \\ 
\chi \\ 
0%
\end{array}%
\right) =m\left( 
\begin{array}{c}
\psi ^{2\dot{2}} \\ 
\psi ^{1\dot{2}} \\ 
\chi \\ 
0%
\end{array}%
\right) ,  \label{B-DKP}
\end{equation}%
respectively, with one zero component, where explicit formulae for the
spinor $p^{A\dot{B}}$ were used, cf. (\ref{4vector-spinor}).

\section{Subsolutions of the Dirac equation}

\subsection{Classical subsolutions of the Dirac equation}

In the $m=0$ case it is possible to obtain two independent equations for
spinors $\xi$, $\eta$\ by application of projection operators $Q_{\pm}=\frac{%
1}{2}\left( 1\pm\gamma^{5}\right) $ to Eq.(\ref{Dirac1}) since $\gamma^{5}%
\overset{df}{=}-i\gamma^{0}\gamma^{1}\gamma^{2}\gamma^{3}$ anticommutes with 
$\gamma^{\mu}p_{\mu}$: 
\begin{equation}
Q_{\pm}\gamma^{\mu}p_{\mu}\Psi=\gamma^{\mu}p_{\mu}\left( Q_{\mp}\Psi\right)
=0.  \label{DiracNeutrino}
\end{equation}
In the spinor representation of the Dirac matrices \cite{Berestetskii1971}
we have $\gamma^{5}=\ \mathrm{diag\,}\left( -1,-1,1,1\right) $ and thus $%
Q_{-}\Psi=\left( 
\begin{array}{c}
\xi \\ 
0%
\end{array}
\right) $, $Q_{+}\Psi=\left( 
\begin{array}{c}
0 \\ 
\eta%
\end{array}
\right) $ and separate equations for $\xi$, $\eta$ follow: 
\begin{subequations}
\label{WEYL}
\begin{align}
\left( p^{0}+\overrightarrow{\sigma}\cdot\overrightarrow{p}\right) \eta & =0,
\label{Weyl1} \\
\left( p^{0}-\overrightarrow{\sigma}\cdot\overrightarrow{p}\right) \xi & =0,
\label{Weyl2}
\end{align}
where $\overset{\rightarrow}{\sigma}$ denotes the vector built of the Pauli
matrices. Equations (\ref{WEYL}) are known as the Weyl equations and are
used to describe massless left-handed and right-handed neutrinos. However,
since the experimentally established phenomenon of neutrino oscillations
requires non-zero neutrino masses, theory of massive neutrinos, which can be
based on the Dirac equation, is necessary \cite{Zralek1997,
Perkins2000,Fukugita2003}. Alternatively, a modification of the Dirac or
Weyl equation, called the Majorana equation, is thought to apply to
neutrinos. According to Majorana theory neutrino and antineutrino are
identical and neutral \cite{Majorana1937}. Although the Majorana equations
can be introduced without any reference to the Dirac equation they are
subsolutions of the Dirac equation \cite{Zralek1997}.

Indeed, demanding in (\ref{Dirac1}) that $\Psi=\mathcal{C}\Psi$ where $%
\mathcal{C}$ is the charge conjugation operator, $\mathcal{C}\Psi=i\gamma
^{2}\Psi^{\ast}$, we obtain in the spinor representation $%
\xi=-i\sigma^{2}\eta^{\ast}$, $\eta=i\sigma^{2}\xi^{\ast}$and the Dirac
equation (\ref{Dirac1}) reduces to two separate Majorana equations for
two-component spinors: 
\end{subequations}
\begin{subequations}
\label{MAJORANA}
\begin{align}
\left( p^{0}+\overrightarrow{\sigma}\cdot\overrightarrow{p}\right) \eta &
=-im\sigma^{2}\eta^{\ast},  \label{Majorana1} \\
\left( p^{0}-\overrightarrow{\sigma}\cdot\overrightarrow{p}\right) \xi &
=+im\sigma^{2}\xi^{\ast}.  \label{Majorana2}
\end{align}

It follows from the condition $\Psi =\mathcal{C}\Psi $ that Majorana
particle has zero charge built-in condition. The problem whether neutrinos
are described by the Dirac equation or the Majorana equations is still open 
\cite{Zralek1997, Perkins2000, Fukugita2003}.

Let us note that the Dirac equations (\ref{Dirac1}) in the spinor
representation of the $\gamma ^{\mu }$ matrices can be also separated in
form of second-order equations: 
\end{subequations}
\begin{eqnarray}
\left( p^{0}+\overrightarrow{\sigma }\cdot \overrightarrow{p}\right) \left(
p^{0}-\overrightarrow{\sigma }\cdot \overrightarrow{p}\right) \xi
&=&m^{2}\xi ,  \label{Dirac2a} \\
\left( p^{0}-\overrightarrow{\sigma }\cdot \overrightarrow{p}\right) \left(
p^{0}+\overrightarrow{\sigma }\cdot \overrightarrow{p}\right) \eta
&=&m^{2}\eta .  \label{Dirac2b}
\end{eqnarray}%
Such equations were used by Feynman and Gell-Mann to describe weak decays in
terms of two-component spinors \cite{Feynman1958}.

\subsection{Other massive subsolutions of the free Dirac equation}

The free Dirac equation (\ref{Dirac1}) in the spinor representation of $%
\gamma $ matrices reads:

\begin{equation}
\left. 
\begin{array}{r}
\left( p^{0}+p^{3}\right) \eta _{\dot{1}}+\left( p^{1}-ip^{2}\right) \eta _{%
\dot{2}}=m\xi ^{1} \\ 
\left( p^{1}+ip^{2}\right) \eta _{\dot{1}}+\left( p^{0}-p^{3}\right) \eta _{%
\dot{2}}=m\xi ^{2} \\ 
\left( p^{0}-p^{3}\right) \xi ^{1}+\left( -p^{1}+ip^{2}\right) \xi
^{2}=m\eta _{\dot{1}} \\ 
\left( -p^{1}-ip^{2}\right) \xi ^{1}+\left( p^{0}+p^{3}\right) \xi
^{2}=m\eta _{\dot{2}}%
\end{array}%
\right\} ,  \label{Dirac2}
\end{equation}%
with $\Psi =\left( \xi ^{1},\xi ^{2},\eta _{\dot{1}},\eta _{\dot{2}}\right)
^{T}$ \cite{Berestetskii1971} (see also \cite{MTW1973,Corson1953} for full
exposition of spinor formalism).

In this Subsection we shall investigate other possibilities of finding
subsolutions of the Dirac equation in the setting of first-order equations.
For $m\neq 0$ we can define new quantities: 
\begin{subequations}
\label{DEF1}
\begin{align}
\left( p^{0}+p^{3}\right) \eta _{\dot{1}}& =m\xi _{(1)}^{1},\quad \left(
p^{1}-ip^{2}\right) \eta _{\dot{2}}=m\xi _{(2)}^{1},  \label{def1} \\
\left( p^{1}+ip^{2}\right) \eta _{\dot{1}}& =m\xi _{(1)}^{2},\quad \left(
p^{0}-p^{3}\right) \eta _{\dot{2}}=m\xi _{(2)}^{2},  \label{def2}
\end{align}

where we have: 
\end{subequations}
\begin{subequations}
\label{DEF2}
\begin{align}
\xi_{(1)}^{1}+\xi_{(2)}^{1} & =\xi^{1},  \label{def3} \\
\xi_{(1)}^{2}+\xi_{(2)}^{2} & =\xi^{2}.  \label{def4}
\end{align}
In spinor notation $\xi_{(1)}^{1}=\psi_{\dot{1}}^{1\dot{1}}$, $%
\xi_{(2)}^{1}=\psi_{\dot{2}}^{1\dot{2}}$, $\xi_{(1)}^{2}=\psi_{\dot{1}}^{2%
\dot{1}}$, $\xi_{(2)}^{2}=\psi_{\dot{2}}^{2\dot{2}}$.

Equations (\ref{Dirac2}) can be now written as 
\end{subequations}
\begin{equation}
\left. 
\begin{array}{r}
\left( p^{0}+p^{3}\right) \eta _{\dot{1}}=m\xi _{(1)}^{1} \\ 
\left( p^{1}-ip^{2}\right) \eta _{\dot{2}}=m\xi _{(2)}^{1} \\ 
\left( p^{1}+ip^{2}\right) \eta _{\dot{1}}=m\xi _{(1)}^{2} \\ 
\left( p^{0}-p^{3}\right) \eta _{\dot{2}}=m\xi _{(2)}^{2} \\ 
\left( p^{0}-p^{3}\right) \left( \xi _{(1)}^{1}+\xi _{(2)}^{1}\right)
+\left( -p^{1}+ip^{2}\right) \left( \xi _{(1)}^{2}+\xi _{(2)}^{2}\right)
=m\eta _{\dot{1}} \\ 
\left( -p^{1}-ip^{2}\right) \left( \xi _{(1)}^{1}+\xi _{(2)}^{1}\right)
+\left( p^{0}+p^{3}\right) \left( \xi _{(1)}^{2}+\xi _{(2)}^{2}\right)
=m\eta _{\dot{2}}%
\end{array}%
\right\}  \label{Dirac3}
\end{equation}

It follows from Eqs.(\ref{DEF1}) that the following identities hold: 
\begin{subequations}
\label{ID2}
\begin{align}
\left( p^{1}+ip^{2}\right) \xi _{(1)}^{1}& =\left( p^{0}+p^{3}\right) \xi
_{(1)}^{2},  \label{id1a} \\
\left( p^{0}-p^{3}\right) \xi _{(2)}^{1}& =\left( p^{1}-ip^{2}\right) \xi
_{(2)}^{2}.  \label{id2a}
\end{align}%
Taking into account the identities (\ref{ID2}) we can finally write
equations (\ref{Dirac3}) as a system of the following two equations:

\end{subequations}
\begin{equation}
\left. 
\begin{array}{r}
\left( p^{0}+p^{3}\right) \eta_{\dot{1}}=m\xi_{(1)}^{1} \\ 
\left( p^{1}+ip^{2}\right) \eta_{\dot{1}}=m\xi_{(1)}^{2} \\ 
\left( p^{0}-p^{3}\right) \xi_{(1)}^{1}+\left( -p^{1}+ip^{2}\right)
\xi_{(1)}^{2}=m\eta_{\dot{1}}%
\end{array}
\right\} ,  \label{constituent1}
\end{equation}

\begin{equation}
\left. 
\begin{array}{r}
\left( p^{1}-ip^{2}\right) \eta_{\dot{2}}=m\xi_{(2)}^{1} \\ 
\left( p^{0}-p^{3}\right) \eta_{\dot{2}}=m\xi_{(2)}^{2} \\ 
\left( -p^{1}-ip^{2}\right) \xi_{(2)}^{1}+\left( p^{0}+p^{3}\right)
\xi_{(2)}^{2}=m\eta_{\dot{2}}%
\end{array}
\right\} .  \label{constituent2}
\end{equation}

Due to the identities (\ref{ID2}) equations (\ref{constituent1}), (\ref%
{constituent2}) can be cast into form:%
\begin{equation}
\left( 
\begin{array}{cccc}
0 & 0 & p^{0}+p^{3} & p^{1}-ip^{2} \\ 
0 & 0 & p^{1}+ip^{2} & p^{0}-p^{3} \\ 
p^{0}-p^{3} & -p^{1}+ip^{2} & 0 & 0 \\ 
-p^{1}-ip^{2} & p^{0}+p^{3} & 0 & 0%
\end{array}%
\right) \left( 
\begin{array}{c}
\xi _{(1)}^{1} \\ 
\xi _{(1)}^{2} \\ 
\eta _{\dot{1}} \\ 
0%
\end{array}%
\right) =m\left( 
\begin{array}{c}
\xi _{(1)}^{1} \\ 
\xi _{(1)}^{2} \\ 
\eta _{\dot{1}} \\ 
0%
\end{array}%
\right) ,  \label{A-D}
\end{equation}%
\begin{equation}
\left( 
\begin{array}{cccc}
0 & 0 & p^{0}-p^{3} & p^{1}+ip^{2} \\ 
0 & 0 & p^{1}-ip^{2} & p^{0}+p^{3} \\ 
p^{0}+p^{3} & -p^{1}-ip^{2} & 0 & 0 \\ 
-p^{1}+ip^{2} & p^{0}-p^{3} & 0 & 0%
\end{array}%
\right) \left( 
\begin{array}{c}
\xi _{(2)}^{2} \\ 
\xi _{(2)}^{1} \\ 
\eta _{\dot{2}} \\ 
0%
\end{array}%
\right) =m\left( 
\begin{array}{c}
\xi _{(2)}^{2} \\ 
\xi _{(2)}^{1} \\ 
\eta _{\dot{2}} \\ 
0%
\end{array}%
\right) .  \label{B-D}
\end{equation}

\section{Supersymmetric equations and their symmetries}

We shall now interpret the subsolutions equations (\ref{A-DKP}), (\ref{B-DKP}%
) and (\ref{A-D}), (\ref{B-D})., First of all, we note that pairs of
equations (\ref{A-DKP}), (\ref{B-DKP}) and (\ref{A-D}), (\ref{B-D}) are
identical in form but have vector and spinor solutions, respectively. We
shall thus refer to these equations as supersymmetric equations. We have
demonstrated that equations (\ref{A-D}) and (\ref{B-D}) are Lorentz
covariant \cite{Okninski2007} and that (\ref{A-DKP}), (\ref{B-DKP}) are
charge conjugated one to another \cite{Okninski2004}.

Let us consider Eqs.(\ref{A-DKP}), (\ref{A-D}). They can be written as:%
\begin{equation}
\gamma ^{\mu }p_{\mu }P_{4}\Psi =mP_{4}\Psi ,  \label{SUSY1}
\end{equation}%
where $P_{4}$ is the projection operator, $P_{4}=$ \textrm{diag }$\left(
1,1,1,0\right) $ and spinor representation of the Dirac matrices.
Incidentally, there are other projection operators which lead to analogous
three component equations, $P_{1}=$\textrm{diag}$\left( 0,1,1,1\right) $, $%
P_{2}=$\textrm{diag}$\left( 1,0,1,1\right) $, $P_{3}=$ \textrm{diag}$\left(
1,1,0,1\right) $ but we shall need only the operator $P_{4}$. Acting from
the left on (\ref{SUSY1}) with $P_{4}$ and $\left( 1-P_{4}\right) $ we
obtain two equations: 
\begin{subequations}
\begin{align}
P_{4}\left( \gamma ^{\mu }p_{\mu }\right) P_{4}\Psi & =mP_{4}\Psi ,
\label{SUSY2a} \\
\left( 1-P_{4}\right) \left( \gamma ^{\mu }p_{\mu }\right) P_{4}\Psi & =0.
\label{SUSY2b}
\end{align}%
In the spinor representation of $\gamma ^{\mu }$ matrices Eq.(\ref{SUSY2a})
is equivalent to (\ref{constituent1}) while Eq.(\ref{SUSY2b}) is equivalent
to the identity (\ref{id1a}). Now the projection operator can be written as $%
P_{4}=\frac{1}{4}\left( 3\mathbf{+}\gamma ^{5}-\gamma ^{0}\gamma
^{3}+i\gamma ^{1}\gamma ^{2}\right) $ (and similar formulae can be given for
other projection operators $P_{1},P_{2},P_{3}$, see \cite{Corson1953} where
another convention for $\gamma ^{\mu }$ matrices was however used). It thus
follows that the supersymmetric equation (\ref{SUSY1}) is now given
representation independent form.

\section{Discussion}

We have shown that subsolutions of the Dirac equation as well as of the DKP
equations for spin $0$\ (similar subsolutions arise in the DKP theory for
spin $1$ \cite{Okninski2003}) obey the Dirac equation with built-in
projection operator (\ref{SUSY1}). Therefore, this covariant equation has
bosonic as well as fermionic degrees of freedom and may provide a background
for supersymmetric formalism. Let us note here that interaction can be
incorporated into \textsl{(\ref{SUSY1}}) via minimal action, $p^{\mu
}\rightarrow \pi ^{\mu }=p^{\mu }-eA^{\mu }$, but in the interacting case
Eq.(\ref{SUSY1}) is non-equivalent neither to the Dirac or the DKP equations 
\cite{Okninski2007}.

\end{subequations}

\end{document}